# Understanding microfabricated nanocalorimeter performance and responses to the energy fluxes from low-temperature plasma discharges

Carles Corbella [1,2,a)], Feng Yi [1], and Andrei Kolmakov [3,b)]

[1]Materials Measurement Science Division, MML, NIST, Gaithersburg, MD 20899, USA
[2]Department Chemistry & Biochemistry, University of Maryland, College Park, MD 20742, USA
[3]Nanoscale Device Characterization Division, PML, NIST, Gaithersburg, MD 20899, USA

[a)] Electronic mail: carles.corbellaroca@nist.gov
[b)] Electronic mail: andrei.kolmakov@nist.gov

Plasma diagnostics have a shortage of fast and sensitive calorimetric sensors that can track substrate temperature during plasma-assisted microfabrication. In this work, energy fluxes from argon and oxygen radiofrequency (RF) glow discharges have been probed using a novel nanocalorimeter sensor. The probe consists of an ultrathin $SiN_x$ membrane (100 nm) with a lithographically defined Pt micro-strip (100 nm) that serves as a calibrated resistance thermometer. The sensor temperature can increase from room temperature to several hundred degrees within a second upon exposure to RF plasma, depending on the experiment's geometry and plasma parameters. Such sensitivity and response time are due to the pre-designed reduced heat capacity of the sensor and significantly reduced thermal conductance of the cooling channels. These features enable the sensitive detection of low-energy plasma fluxes on surfaces and their rapid discrimination, as in the case of ion and electron fluxes, by biasing the sensor at negative or positive potentials. These biased nanocalorimeter energy readings have been compared with ion and electron kinetic energy dissipations assessed using a Langmuir probe and retarding field energy analyzer (RFEA). Finally, the robustness of the plasma nanocalorimeter is discussed in terms of its baseline drifts, degradation, and longevity.

# I.  INTRODUCTION

Plasma cleaning, plasma-assisted deposition, and etching are indispensable processes in the semiconductor fabrication industry[1-5]. Unlike the widely used optical and electrical plasma diagnostics metrologies, calorimetric or generally heat flux probes constitute a less conventional, yet evolving metrology that utilizes the thermal effects of radiation and particle fluxes for plasma diagnostics[6-9].

Langmuir and capacitive probes, Faraday cup[10-13], retarding field energy analyzer (RFEA)[14-16], mass spectrometry[17], and optical spectroscopies[18,19], are established diagnostic tools to characterize plasma-assisted microfabrication processes. However, these techniques are not directly suited to track the thermal loads at the substrate level. On the other hand, *in situ* real-time monitoring of heat fluxes with sufficient sensitivity and response time is crucial for assessing the viability and outcome of material processing using plasma discharges[20]. The measured heat fluxes should also provide benchmarks to support fundamental plasma-surface interaction studies[21,22]. Therefore, for scientific and industrial purposes, the development of miniaturized, fast calorimeters capable of efficiently operating during long plasma processes is a compelling milestone. Plasma nanocalorimeter's design and performance constitute the motivation of this article.

Various designs of passive and active calorimetric probes, primarily based on thermocouples or thermopiles, often featuring specific catalytic coatings, have been developed for the selective sensing of plasma reactive species[23-27]. Besides its simplicity and versatility of designs and readouts, such metrology offers excellent repeatability and sensitivity. Current efforts in plasma thermal probes are focused on increasing their sensitivity, selectivity, and shortening response time[28,29]. The latter will allow for expansion of the pressure range[30] and operation during fast plasma transients relevant to low- to medium-frequency pulsed plasma-assisted processes[31,32]. Additionally, efforts are underway to reduce the calibration time of sensors and enhance their attractiveness by incorporating Langmuir probe-like functionalities[33].

Though the catalytically sensitized calorimetric sensors are tuned to selectively measure the fluxes of reactive radicals, the general response of the calorimetric sensor to other plasma energy fluxes remains a practically important research field. The latter has been thoroughly reviewed by Kersten and others[8,34]. In essence, the heat flux from plasma delivered to a surface can be expressed as the sum of the following components[23]:

$$J_{\text{plasma}} = J_i + J_e + J_{\text{ph}} + J_n \qquad (1)$$

where $J_i$ is the energy flux contribution from ions including their kinetic and potential (recombination) energies components, $J_e$ is the energy flux from incoming electrons (including effects of secondary electron emission), $J_{\text{ph}}$ is the radiative flux of plasma-generated photons (including UV and IR partitions), and $J_n$ is the energy flux from neutral species, like energetic neutrals, radicals, and metastable atoms. The latter also includes the heats of adsorption, recombination, and condensation reactions. At steady state, the incoming thermal energy from these heat sources must be balanced by the heat dissipation mechanisms of the substrate via thermal conduction (both through the substrate and the gas phase), with convection and radiation also relevant at elevated pressures and temperatures, respectively[34].

The sensitivity and the range of heat fluxes measured using a calorimetry sensor strongly depend on its design and the selection of active materials. Additionally, the time resolution, and therefore the data acquisition rate, are directly related to the sensor's heat capacity. Thus, it would be advantageous to employ low-thermal-mass sensors to monitor temperature variations induced by different plasma species. These sensors with reduced thermal mass, known as nano-(or micro-)calorimeters, are widely used for fast calorimetric analysis[35]. They essentially consist of a microfabricated, self-supported thin film multilayer suspended on a silicon frame, and they possess unique characteristics such as high temperature sensitivity and very short response times. However, the application of nanocalorimeters for technical plasma diagnostics is rather recent[29], and many questions, such as their relative responses to specific energy fluxes and the longevity of the ultrathin sensor in harsh plasma environments, require further study.

In this work, we test a fast, sensitive nanocalorimetry sensor based on an ultrathin Pt resistor lithographically defined on top of a suspended $SiN_x$ membrane[29,36,37]. Its performance and stability have been characterized using argon and oxygen RF inductively coupled (ICP) model plasmas generated in a custom-made research reactor. Plasma parameters have been adjusted to mimic those used in microfabrication processing. For this, the plasma discharges were pre-characterized using Langmuir probe and retarding field energy analyzer (RFEA). The operation of a remote RF ICP source, together with the option to apply a DC bias to the sensor, allowed for independent control over plasma density and kinetic energy of incident charged particles[1]. The role played by each heat agent has been resolved by tuning the energy and flux of ions and measuring corresponding changes in the temperature readings. Such changes have been correlated to the electric power dissipation mechanism. The article concludes with a discussion of the lifetime of the nanocalorimeter exposed to reactive (oxygen) and non-reactive (argon) plasmas.

## II. EXPERIMENTAL

### A. Plasma reactor and diagnostics

#### 1. Vacuum system

The plasma nanocalorimetry experiments were performed in a dedicated reactor equipped with a set of plasma diagnostics tools (Figure 1). The core consists of a vacuum cross chamber equipped with a magnetic manipulator for horizontal and rotational movements. The electrically grounded sensor holder is mounted on the manipulator using an L-shaped bracket. The nanocalorimeter sensor die is electrically and thermally decoupled from the grounded holder, allowing for external DC biasing. A remote RF ICP discharge source (13.56 MHz) operating between 10 W and 100 W is installed on the top flange. The generated plasma propagates to the holder position through a vertical quartz tube approximately 20 cm long and with an inner diameter of 22 mm. Its lower end is capped by an electrically grounded, showerhead Al electrode foil with a set of punctured small holes arranged to improve plasma uniformity in front of the sample. The gap between the Al shower cap and the sensor surface is 12 mm, and it is filled with a plasma plume at RF power around 50 W and higher. The plasma flux can be conveniently interrupted by means of a manual shutter. The reactor is pumped through the throttle valve using a turbomolecular pump (0.07 m$^3$/s) backed by two mechanical pumps in series. The base pressure is lower than 10$^{-3}$ Pa. Ar and O$_2$ gases of ultra-high purity (0.99999 mol/mol) were used in the present study. The plasma pressure is measured using a Baratron gauge and can be finely controlled by adjusting the gas delivery rate and pumping speed.

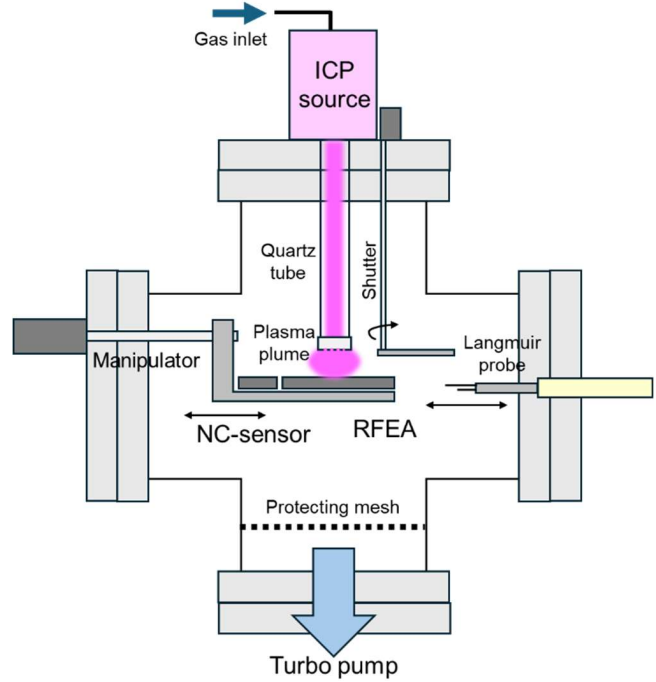

FIG. 1. Schematic frontal view cross-section of the plasma reactor assembled to test the nanocalorimeter (NC). Main components and plasma diagnostics systems are indicated. ICP: inductively coupled plasma; RFEA: retarding field energy analyzer.

## 2. Plasma characterization

The plasma monitor set consists of the following elements:

- *Langmuir probe:* Plasma density, electron temperature, and plasma potential, as well as the electron energy distribution function (EEDF), were measured using an RF-filtered Langmuir probe along with a compensation electrode to avoid I-V curve distortions near plasma potential. The probe tip consists of a 10 mm long and 0.4 mm diameter tantalum wire, and it can be moved along the same axis of the magnetically driven holder.

- *Retarding field energy analyzer (RFEA):* The L-shaped holder described above supports a nanocalorimeter enclosure along with a 10 cm diameter RFEA housing. A compact stack of fine parallel grids, with a total distance between the upper grid and the collector of approximately 0.5 mm, minimizes the effect of ion scattering due to internal collisions at elevated working pressures. The ion current versus retarding voltage is measured and converted into an ion energy distribution (IED). Ion energies of up to 2000 eV can be collected by DC biasing of the RFEA housing.

The nanocalorimetry experiments reported hereafter were performed within the range of 1 Pa to 10 Pa and 80 W of RF power to remain within the range of plasma parameters (e.g., density, ion flux, electrons/ions energies) relevant to those used in industrial ICP cleaning and etching applications (Table I)[1,38].

TABLE I. Plasma parameters for Ar and O$_2$ ICP discharges measured using Langmuir probe and RFEA: RF power ($P_{RF}$), pressure ($p$), plasma density ($n$), electron temperature ($T_e$), plasma potential ($V_p$), peak ion energy ($E_i$), and ion flux ($j_i$). Typical values used for plasma processing in microfabrication laboratories are listed for comparison[1,38].

|  | Argon | Oxygen | Microfab. |
|---|---|---|---|
| $P_{RF}$ (W) | 10 to 80 | 60 to 80 | >100 |
| $p$ (Pa) | 1.3 to 6.5 | 6.5 | 0.1 to 150 |
| $n$ (m$^{-3}$) | $1\times10^{15}$ to $5\times10^{16}$ | $3\times10^{15}$ to $5\times10^{15}$ | $10^{14}$ to $10^{19}$ |
| $T_e$ (eV) | 2 to 5 | 4 to 5 | 1 to 7 |
| $V_p$ (V) | 20 to 30 | 35 to 40 | 20 to 40 |
| $E_i$ (eV) | 20 to 130 | 35 to 140 | 20 to 2000 |
| $j_i$ (A/m$^2$) | 1 to 30 | 15 to 25 | 0.01 to 500 |

### B. Plasma nanocalorimetry

#### 1. Nanocalorimeter sensor

The nanocalorimeter assembly used in this study is a version of the system reported elsewhere[29]. Pt resistor microstrips were defined on low-stress free-standing ultrathin SiN$_x$ membranes (2 mm × 7.25 mm area) on silicon die frames by means of lithography processes (Figure 2). In short, a polished, 500 µm thick silicon wafer was first coated with 100 nm SiN$_x$ films on both sides. Subsequently, one side was coated with a 100 nm Pt layer using 10 nm Ta as an adhesion layer. Then, Pt microstrips and electrode pads were lithographically defined. The Si wafer is back-etched with KOH up to the SiNx interface with a Pt structure on top. In a last fabrication step, the robustness of the nanocalorimeter sensor is enhanced by air annealing of the die[36]. The final structure consists of a SiNx membrane facing the plasma discharge, while the opposite surface, featuring Pt resistors and contact pads, rests on a printed circuit board (PCB) via spring-loaded pins (Figure 2).

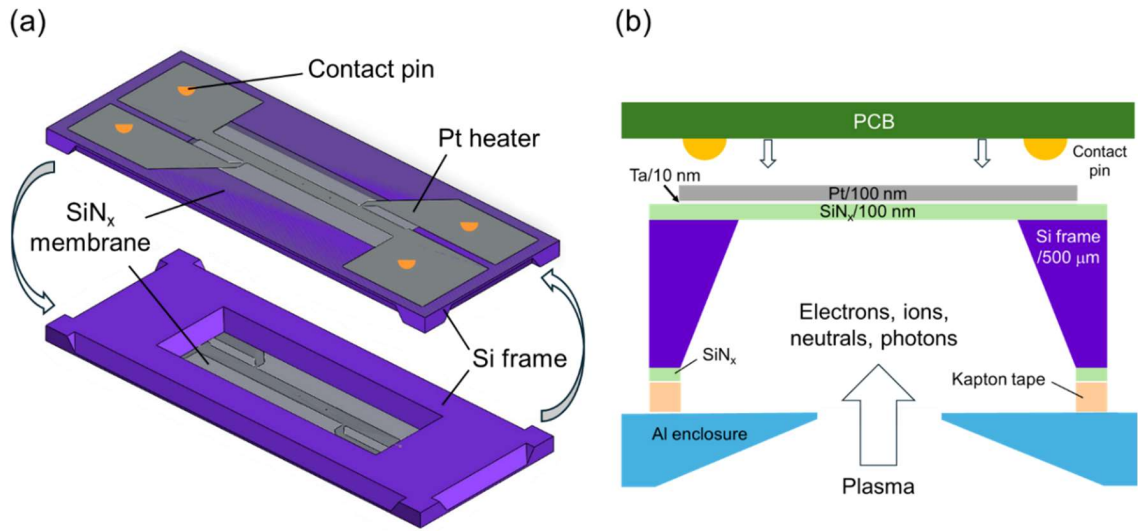

FIG. 2. (a) 3D schematics of the nanocalorimeter: a silicon die with a suspended membrane with evaporated Pt resistive sensor and contact pads. (b) Schematic cross-section of the nanocalorimeter (not to scale) showing the layered structure of the free-standing Pt/Ta/SiN$_x$ thin film structure and connection with the printed circuit board (PCB).

The Pt resistive sensor structure is designed for 4-probe measurements. A temperature calibration of the Pt resistor is performed in a separate furnace prior to experiments. Accordingly, the average temperature coefficient of resistance (TCR) of the device is 0.025 $\Omega$/K[36]. A heat capacity of approximately $2\times10^{-6}$ J/K for the suspended film stack has been determined via a temperature relaxation method (see *Supplementary Material*)[39]. This experimental value roughly matches the nominal heat capacity derived from material data. The nanocalorimeter sensor chips (5 mm × 13 mm) can be mounted either as a single chip or in pairs (for differential nanocalorimetry) inside an electrically grounded aluminum enclosure. The sensor die is isolated from the enclosure by a 0.1 mm thick Kapton film to reduce the heat dissipation and to allow for the electrical bias of the sample. The sensing area of the device is limited by a rectangular opening in the enclosure (slit area = 0.7 mm × 4.6 mm), which faces the plasma plume from the quartz tube directly.

## 2. Measurement setup

The temporal evolution of the nanocalorimeter resistance in plasma is measured using a computer-controlled digital source meter connected to the PCB in 4-probe configuration (Figure 3). The injected current was set at 1 mA, and the sensed potential drop over the Pt resistor was on the order of 10 mV at room temperature. The measured resistance can then be converted to sensor temperature using a calibration curve. Additionally, a separate DC power supply was used to bias the floating nanocalorimeter, allowing for modulation of the flux and energy of the incident ions and electrons. This setup is used to decouple the contributions of heat fluxes from charged and neutral particles (plus radiation) to the total measured heat flux. The resulting current from plasma is measured with a current meter. Given the high sensitivity of the measurement setup, electrical connections from the sensor enclosure to the vacuum feedthrough require metallic shielding to reduce electric noise induced by plasma.

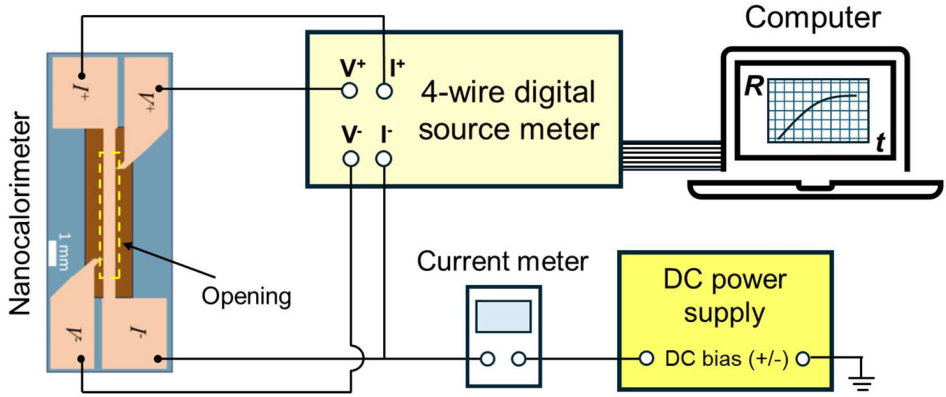

FIG. 3. Setup for computer-controlled 4-probe resistance measurements with option to apply a DC bias on electrically floating sensor. The plasma-exposed area of 0.7 mm × 4.6 mm on the active side (opposite the Pt resistor) is enclosed by a rectangular dashed frame.

## III. RESULTS AND DISCUSSION

### A. Analysis of heating/cooling curves

#### 1. Nanocalorimeter response to the heat flux from plasma

Figure 4 illustrates the evolution of Pt microstrip resistance and its corresponding temperature upon exposure to Ar plasma. The temperature response for both Ar and $O_2$ plasma exposures is provided in the *Supplementary Material*. These measurements were performed using an electrically grounded sensor. The zero-time onset corresponds to the shutter open state and manifests a sub-second response of the Pt sensor to the heat flux from the plasma. Note that the noise level of the measured resistance signal remains constant regardless of whether the shutter is open or closed. The observed oscillations are therefore attributed to noise generated by the current source unit, which translates into self-heating noise of the sensor.

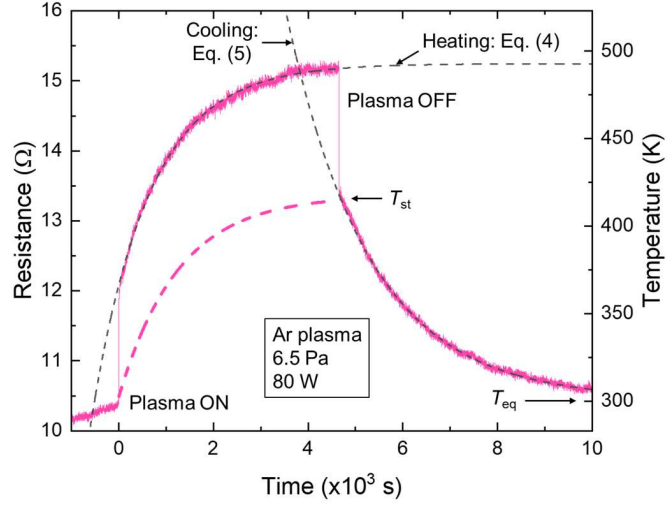

FIG. 4. Temporal evolutions of resistance and temperature of the nanocalorimeter (electrically grounded) when exposed to Ar plasma. Black dashed lines are fitting curves to the exponential heating/cooling laws expressed in Eq. (4) and Eq. (5). The fitted parameters are discussed in the text. The pink dashed line corresponds to the slow temperature increase of the sensor envelope during plasma exposure.

The fast transient is followed by a much slower process of stabilization, reaching temperatures of around 450 K to 500 K, which is achieved after roughly one hour of plasma operation. The latter temperature values corroborate typical gas temperatures reported for ICP discharges[40,41]. In general, the temperature dynamics of the nanocalorimeter sensor, $T_s$, has been described using the classical heat balance equation[23,33]:

$$P_{in} = \dot{H}_s + P_{out} \qquad (2)$$

where $P_{in}$ and $P_{out}$ are the incoming and outgoing heat fluxes (powers) to/from the sensor, respectively. $\dot{H}_s$ is the time derivative of temperature multiplied by the sensor heat capacity, $C_s$:

$$\dot{H}_\text{s} = C_\text{s} \cdot \dot{T}_\text{s} \tag{3}$$

In our design, an initial, rapid temperature increment stage of heating from room temperature occurs as soon as the shutter opens. The temperature then continues to increase gradually. After the near-saturation condition is achieved, the heat flux is interrupted by switching the plasma OFF, and a fast temperature transient, similar in magnitude to the initial heating stage, occurs. Such rapid heating/cooling onsets are attributed to the fast heating/cooling of the membrane-based Pt microstrip by the plasma thermal flux. Note that at saturation $P_\text{in} = P_\text{out}$ and $P_\text{out} = \dot{H}_\text{s}$ when the plasma is switched OFF from Eq. (2). Therefore, the incoming power can be estimated by knowing $C_\text{s}$ and calculating the temperature derivative at the moment when the plasma is OFF. Here, we assume that heat dissipation is solely due to thermal conductance, and that radiative as well as heat losses through the gas phase are negligible at the used temperature and pressure ranges[33]. Thus, estimated incoming power values to the sensor for Ar and $O_2$ plasmas are approximately 2.05 mW and 1.30 mW, and the corresponding heat fluxes corrected for the opening (slit) area are therefore 640 W/m$^2$ and 400 W/m$^2$, respectively. Note that the time interval used to estimate $\dot{T}_\text{s}$ is 50 ms, which is comparable to the diffusive molecular transport timescale of plasma species, and it corroborates nicely with the response time of reactive H radicals measured in a previous work with a similar reactor setup[29]. The estimated plasma heat fluxes are two orders of magnitude higher than 10 µW dissipated by the 1 mA probing current, thereby justifying that sensor self-heating is negligible in the discussion.

## 2. Long-term substrate temperature drift

As plasma exposure continues, after a sub-second sensor response, the temperature curves tend to saturate slowly due to the establishment of thermal equilibrium between incoming and outgoing heat fluxes, resulting in $P_{in} = P_{out}$. After the thermal flux is halted, the nanocalorimeter undergoes a rapid sub-second cooling transient, followed by a slower cooling phase. These slow heating and cooling evolutions are due to the thermal response of the entire sensor assembly (including the Al enclosure). During the gradual heating stage, sensor temperature follows an exponential law, assuming that heat dissipates predominantly via thermal conduction[33]:

$$T_s(t) = \left(T_{eq} + \frac{P_{in}}{a}\right) - \left(\frac{P_{in}}{a}\right)\exp\left(-\frac{a}{C_s^*}t\right) \qquad (4)$$

Here, sensor temperature depends on the environment (equilibrium) temperature, $T_{eq}$; the incident plasma power onto the sensor, $P_{in}$; a heat transfer coefficient, $a$, and an effective heat capacity, $C_s^*$. During the cooling (plasma OFF) state, we have accordingly:

$$T_s(t) = T_{eq} + (T_{st} - T_{eq})\exp\left(-\frac{a}{C_s^*}t\right) \qquad (5)$$

which includes the temperature when the cooling process starts, $T_{st}$. Note that, different from the "instant" $C_s$, the $C_s^*$ In Eq. (4) and Eq. (5) is an effective "long-term" thermal capacitance of the sensor, which includes the contribution from the significantly more massive sensor frame.

The fitted $C_s^*$ value is 0.020 J/K, four orders of magnitude bigger than $C_s$ ($2\times10^{-6}$ J/K), and it is similar to the approximate thermal mass of the silicon die after considering specific heat and density of crystalline silicon. In addition, the corresponding $a$ values are $1.6\times10^{-5}$ W/K and $1.1\times10^{-5}$ W/K for the heating and cooling phases, respectively. They represent combined energy losses into the gas phase and Si frame. The higher heat transfer coefficient in the heating phase agrees with the higher thermal conductivity expected in a gas discharge (Wiedemann-Franz law)[1].

## B. Influence of different plasma species on heating/cooling curves

### 1. Measurement of ion energy distributions

A discrimination between the different plasma contributions from Eq. (1) is preferred to investigate the major heating mechanisms of the nanocalorimeter. In order to isolate the heating exerted by ion fluxes onto the nanocalorimeter surface, plasma experiments were conducted by DC biasing of the sensor, thus tuning the kinetic energy of ions impinging on the sensor.

Experimental ion energy distribution (IED) curves corresponding to Ar and $O_2$ plasmas, together with measured corresponding total ion fluxes, are shown in Figure 5. As expected, the IED peaks exhibit a finite width due to ion scattering within the presheath, and small-intensity low-energy tails are due to charge-neutral exchange scattering in the plasma sheath. The upshift of the energy peak in IED is due to the application of a negative DC bias voltage. The measured ion energy peak, $E_i$, is proportional to the difference between plasma potential, $V_p$, and applied DC bias, $V_{bias}$:

$$E_i = e(V_p - V_{bias}) \qquad (6)$$

where $e$ is the elementary charge ($1.602 \times 10^{-19}$ C). Here, ionic species in plasma are assumed to be singly ionized. Plasma potentials, measured with the Langmuir probe, are circa 30 V for argon and 40 V for oxygen.

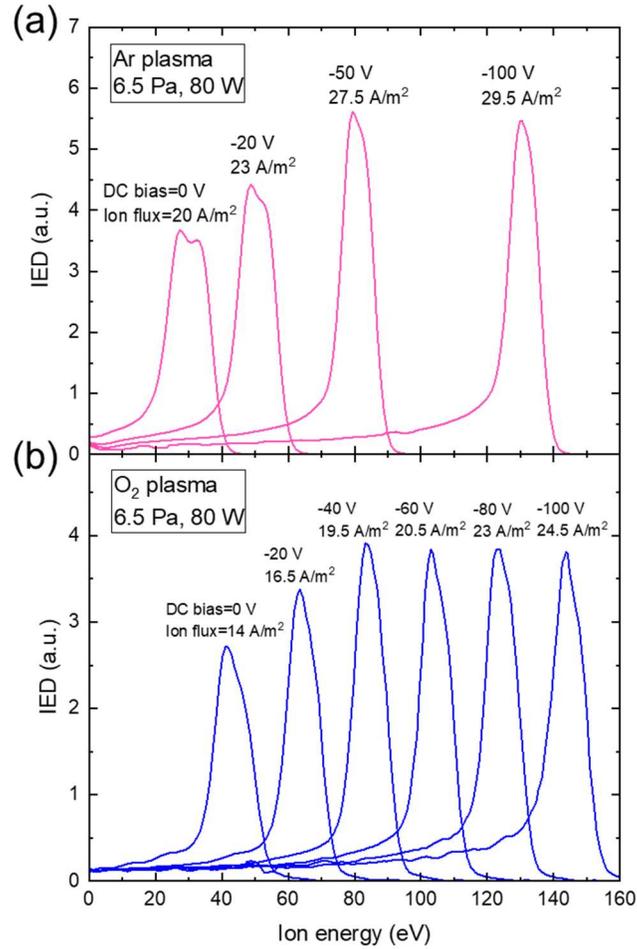

FIG. 5. Ion energy distributions (IED) of (a) Ar and (b) $O_2$ plasmas at DC bias (sensor voltage) scanned from 0 V to -100 V measured by retarding field energy analyzer (RFEA). The measured ion fluxes, equal to the integrated area under each curve, are indicated.

## 2. Contribution from plasma ions

Figure 6a shows the sensor temperature curves acquired under an incremental 10 V increase of negative DC bias up to -100 V. The resultant staircase-shaped heating curves reflect the variation in sensor temperature due to a bias-induced increase in ion flux and energy. At a given DC bias, the measured step in resistance or temperature is more pronounced in Ar plasma than in O$_2$ plasma, likely due to the higher ion current in the former. The higher ion current in the Ar discharge can be attributed to the higher plasma density values (Table I).

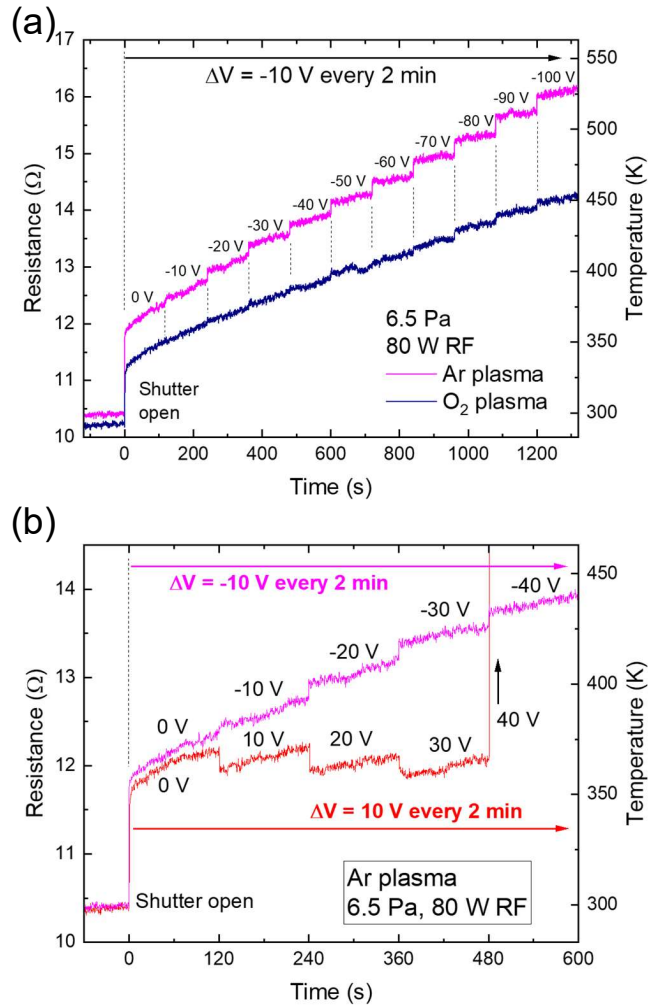

FIG. 6. (a) Staircase-shaped sensor temperature change in Ar and O$_2$ plasmas upon increase of negative DC bias voltage and resultant acceleration of positive ions. (b) Comparison of the heating curves obtained under positive and negative DC biases for the case of Ar discharge. The bias value increases every 120 s in all cases.

Heat flux variations resulting from a reduction in ion kinetic energy have also been recorded. The heating curves corresponding to Ar plasma ions under either positive or negative DC bias applied to the nanocalorimeter sensor are compared in Figure 6b. Their evolutions exhibit mirror-like behaviors: while upward steps in temperature were generated by an increase in negative DC bias, the opposite, i.e., downward steps, were observed in the case of a positive DC bias increase. This "endothermic-like" behavior is consistent with the retarding effect that a positive bias exerts on the kinetic energy and flux of incoming ions, as discussed below. Note that the experiment had to be discontinued after applying bias voltages higher than 40 V due to a dramatic increase in electron current at approximately the plasma potential, which led to sensor overheating, as registered temperatures immediately surpassed 500 K.

### 3. Contributions from plasma electrons, photons, and neutrals

The notorious variations in temperature upon changes in DC bias indicate the crucial effects of ion and electron kinetic energies on the thermal behavior of the nanocalorimeter. This effect can also be used to decouple sensor heating by ions from other plasma species, as discussed for Ar discharges in the following paragraphs.

Discrimination of ion-induced heating is addressed by determining the positive DC bias voltage range in which the thermal impact from energetic plasma ions is minimized, thereby leaving electrons, photons, and neutrals as the primary heating agents. Such retarding DC potential should be sufficiently high to prevent positive ions from reaching the sensor, and yet it should also be below the plasma potential to avoid attracting a high flux of electrons. This balance range, between 0 V and 40 V, can be found by minimizing the temperature in response to increasing DC bias (see the range between 2300 s and 2600 s in Figure 7 as an example). The combined heat flux from plasma photons, energetic neutrals, and electrons can therefore be computed from temperature drop, $\dot{T}_s$, as plasma is switched OFF.

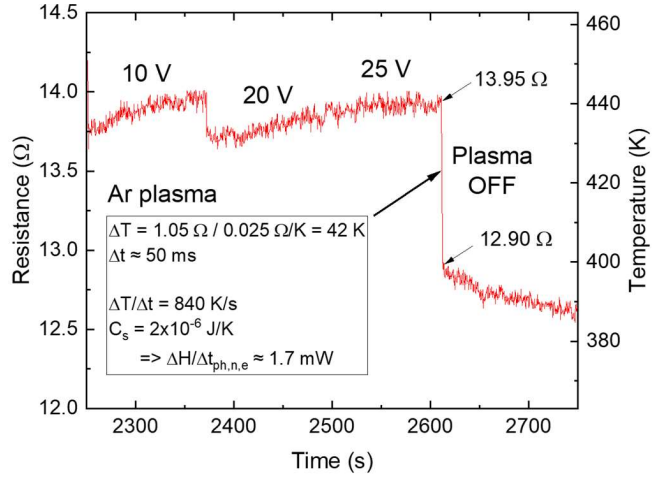

FIG. 7. Determination of the heating power delivered to the sensor by electrons, photons, and neutrals (ions excluded), $\dot{H}_{ph,n,e}$, from Ar ICP discharge at 6.5 Pa and 80 W of RF power. Applied DC bias voltages are indicated.

The application of Eq. (3) yields a minimum of heating power of 1.7 mW or heat flux of 520 W/m² for Ar plasma. The nanocalorimeter sensitivity with the currently used detection scheme is approximately 10 % of this flux. Consistently, this parameter is lower than 2.05 mW, previously derived at 0 V (grounded). The effective heating input from photons, neutrals, and electrons will be compared to the heat flux generated solely by plasma ions in the next section.

## C. Details on ion-induced heating

### 1. Measurement of ion current

Here, we analyze two related measurands in more detail and discuss the correlation between them. In particular, we compare the ion-induced sensor heating power with DC bias-controlled ion kinetic energy flux ($j_i \cdot E_i$) dissipated in the nanocalorimeter. For this purpose, a similar approach to that developed by Bornholdt and Kersten[33] was employed, measuring the DC bias-dependent total current of charge particles at the nanocalorimeter Pt sensor as shown in Figure 3. Although the $SiN_x$ layer over the Pt sensor has, in principle, low electrical conductivity, it appears to be conductive enough to neutralize the incoming microampere-level current of ions or electrons. This is presumably due to the small thickness and non-stoichiometry of the $SiN_x$ layer, as well as plasma-induced photoconductivity in the membrane. However, we cannot exclude the formation of a small (on the order of Volts) steady state potential at the surface of the $SiN_x$ membrane.

Figure 8 shows I-V curves recorded on Pt strip by sweeping DC bias voltage between -100 V and 100 V. The Ar curve qualitatively resembles a typical planar Langmuir probe I-V characteristics[33]. Consistent with the Langmuir probe measurements above, plasma potential shift is observed near 30 V (see inset in Figure 8), although small discrepancies in plasma parameters are admissible due to the distinct probe sizes and geometries (cylindrical probe vs. planar probe). A different scenario is registered in case of $O_2$ plasma, where the current at positive DC voltages is even smaller than the positive ion currents recorded in the negative voltage region.

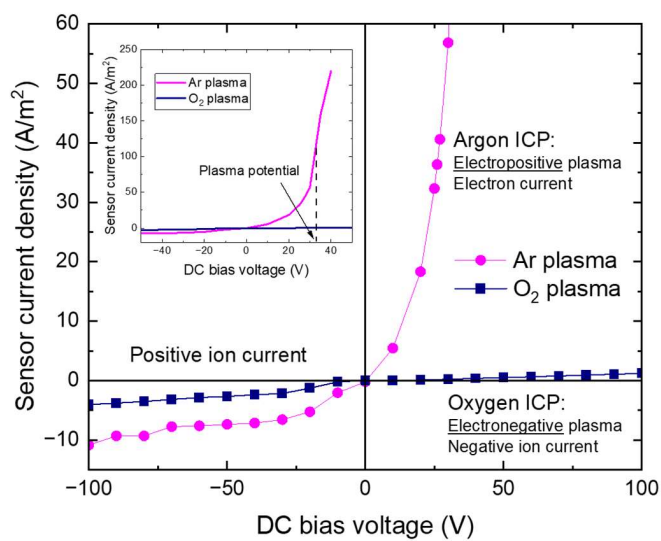

FIG. 8. Current-voltage characteristics measured through the nanocalorimeter resistor sensor upon exposure to Ar and $O_2$ plasmas (6.5 Pa, 80 W). Dominant currents and role played by plasma electronegativity for each gas are mentioned. Inset: I-V curve with a rescaled Y-axis indicating the approximate plasma potential for Ar discharge.

- *Negative DC bias branch:* Ion current measured to be higher for Ar plasma than for $O_2$ plasma under the same conditions, presumably due to the higher plasma density in Ar discharge. Maximal ion current densities of 11 $A/m^2$ and 4 $A/m^2$ are recorded for Ar and $O_2$ plasmas, respectively. These values appear to be lower than ion currents measured with RFEA by a factor 5 (see Figure 5), probably due to the non-optimized current collection geometry of the nanocalorimeter setup and to the limited electrical conductance of the $SiN_x$ layer that can lead to uncontrolled surface charging.
- *Positive DC bias branch:* Measurements with Ar plasma show that the preferential attraction of electrons at positive bias causes a drastic current increase at values around plasma potential and higher (Figure 6b). High electron currents recorded at DC potentials higher than 30 V led to sensor overheating and eventual failure. On the other hand, the electronegative $O_2$ plasma exhibited a gradual increase up to only 1.2 A/m² at a 100 V DC positive bias. Consequently, different from the Ar case, sensor heating measured in oxygen plasma at positive voltages is relatively low (not shown here). This reduced current is attributed to the dominant density of $O^-$ ions over electrons, a characteristic common to electronegative plasmas.

### 2. *Heat flux due to ion kinetic energy dissipation*

The possibility of measuring temperature increments upon incremental changes in negative DC bias values (Figure 6) allows for analysis of the ion kinetic energy contribution to the total heat flux to the sensor.

Heat flux delivered in each -10 V DC bias incremental step [Eq. (3)] is due to ion kinetic energy dissipated to the sensor. The ion kinetic energy dissipation flux (IKEDF) is defined as:

$$P = j_i \cdot E_i \tag{7}$$

where $j_i$ is the ion current density, and $E_i$ is ion kinetic energy, approximated for simplicity as the peak ion energy from IED (Figure 5).

For the sake of brevity, only Ar plasma is considered here due to the stronger signal compared to $O_2$ plasma. Note that the ion kinetic energy flux defined here should not be confused with Joule heating by electron current flowing through the sensor Pt strip or membrane, which is several orders of magnitude lower.

Figure 9a shows the sensor temperature evolution during negative DC bias excursions from 0 V to -100 V (heating-up steps) and back from -100 V to 0 V (cooling-down steps). The separately measured heating curve baseline at 0 V bias (see also Figure 4) is shown in pink and represents the heating profile of the holder without ion acceleration by an external DC bias. By subtracting the thermal drift at zero bias from the recorded temperature curve, one can obtain the temperature profile due to the influx of ion kinetic energy (Figure 9b).

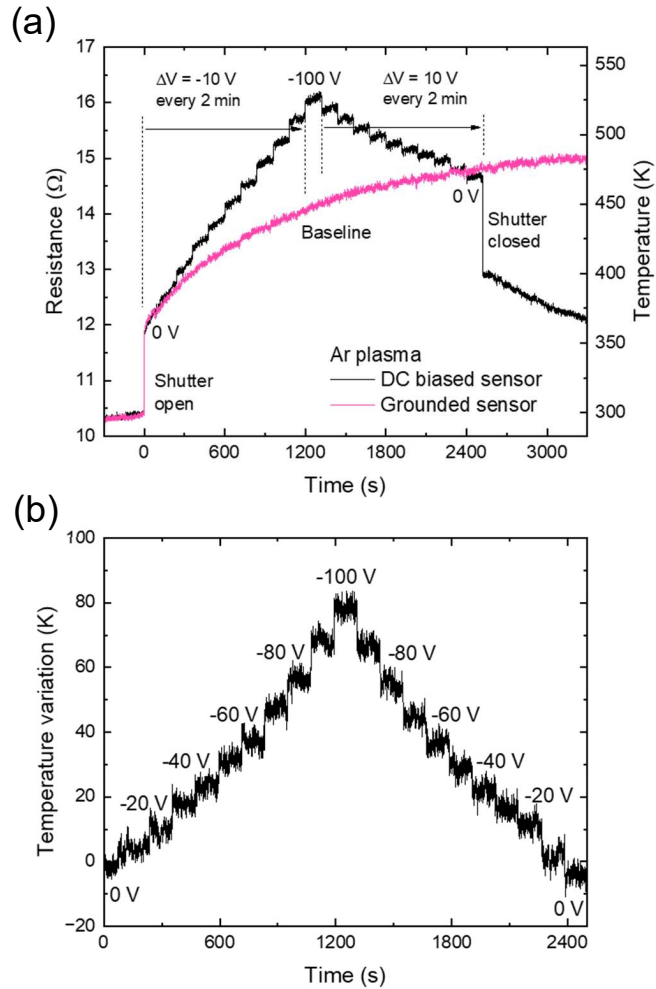

FIG. 9. (a) Symmetric DC bias staircase-shaped temperature evolution together with baseline defined by heating curve at 0 V (Ar plasma, 6.5 Pa, 80 W). (b) The thermal profile resulting from the contribution of ion kinetic energy, after subtracting the sensor holder temperature drift.

Here, we define the total heat flux as the sum of ion heat flux (IHF) and neutral heat flux (NHF), which represent the thermal flux contributions from ions and neutral species, respectively. IHF is calculated from temperature variations shown in Figure 9b according to Eq. (3), while NHF is equal to 520 W/m$^2$ in our case (see sec. III.B.3). Figure 10 shows IHF evolving as a function of DC bias applied to the nanocalorimeter. In addition, IKEDF calculated from Eq. (7) is depicted as a function of DC bias for comparison. Both IHF and IKEDF increase monotonically from 0 V to -100 V, showing an excellent agreement within the explored range.

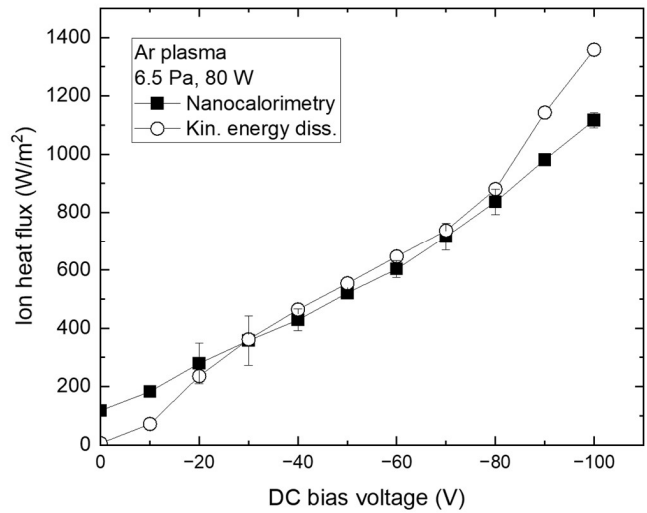

F$_{IG}$. 10. Ion heat flux (IHF, solid squares) at different DC bias measured via nanocalorimetry from Ar plasma data in Figure 9, along with ion kinetic energy dissipation flux (IKEDF, open circles). Nanocalorimetry data correspond to the mean values of heating-IHF and cooling-IHF at each DC bias voltage. Error bars in nanocalorimetry data are defined as the difference between heating and cooling IHF.

The total heat flux interval ranges between 600 W/m² to 1700 W/m². The explored interval matches reasonably well with other related plasma experiments using calorimetric probes[23]. The percentage of thermal flux delivered by ions from the total heat flux is only 15 % at 0 V. This ratio increases steadily up to 70 % at -100 V, thus proving that sensor heating is dominated by ion bombardment at moderate DC bias voltages.

The agreement between ion heat fluxes measured via nanocalorimetry and ion kinetic energy dissipation supports the hypothesis that energetic ion collisions drive sensor heating from plasma, as pointed out elsewhere[7,23].

### D. In-plasma SiN$_x$ membrane lifetime considerations

#### 1. Sputtering rate

Here, we discuss the nanocalorimeter lifetime based on expected sputtering rates under different operation conditions and report the longevity of a sensor exposed to a low energy plasma. Theoretical estimations of the etching rate of the 100 nm membrane and expected lifetime for the nanocalorimeter exposed to Ar and O$_2$ plasmas (6.5 Pa, 80 W) and biased at different voltages are depicted in Figure 11. Silicon is considered a target material in estimations due to the larger number of experimental data available and to the reduced nitrogen concentration in the sub-stoichiometric SiN$_x$ membrane. The sputtering rate is calculated as:

$$R = \frac{M \cdot Y \cdot j_i}{e \cdot \rho} \tag{8}$$

which includes experimental sputter yields of silicon (target) by argon and oxygen ions, $Y$ [42]; Si mass density, $\rho$ (2900 kg/m³); Si atom mass, $M$ (28.1 u), and ion current density measured at the sensor, $j_i$. Nominal lifetime of the membrane at a given ion energy is defined as the time necessary to etch 100 nm of Si target. Ion penetration depth has been neglected here as it is limited to a few nanometers according to TRIM (Transport of Ions in Matter) simulations of Si bombardment with argon and oxygen ions[43]. This framework is limited to non-reactive sputtering, excluding chemical modifications of the target surface. The real scenario of reactive sputtering is much more complex and would require a dedicated study.

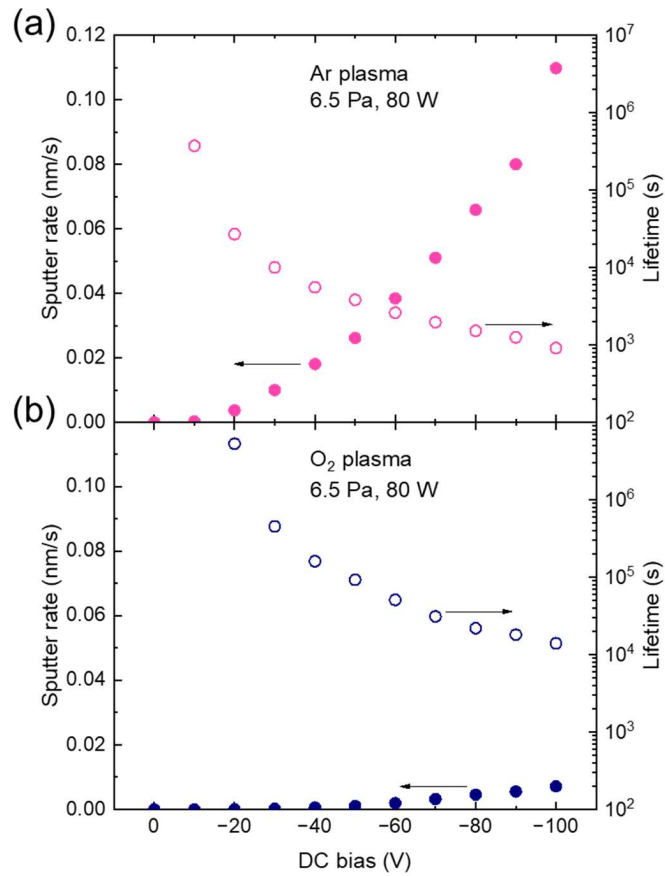

FIG. 11. Calculated sputter rates [Eq. (8)] and expected lifetimes of 100 nm $SiN_x$ membranes exposed to (a) Ar and (b) $O_2$ plasmas. A non-reactive sputtering theory was used[42].

Sputtering rates increase up to 0.11 nm/s for Ar plasma and up to 0.007 nm/s for $O_2$ plasma with ion kinetic energy controlled by DC bias. The significantly higher sputter yields and ion currents in the argon discharge result in Ar sputter rates overcoming oxygen rates by one order of magnitude. The expected membrane lifetime is therefore much higher in the case of oxygen plasma exposure, reaching almost four hours at -100 V DC bias on the sensor.

### 2. *Membrane longevity*

To verify these predictions, a longevity test was conducted on a nanocalorimeter sensor using a custom-made parallel-plate cell in a capacitively coupled plasma (CCP) configuration. This configuration allows for continuous operation for many hours, providing a better representation of real-world processing conditions compared to a quasi-remote plasma source setup. The grounded sensor with holder enclosure was facing a 50 mm diameter plate acting as an active electrode. The electrode gap was 13 mm. RF power of 20 W can be continuously fed into an Ar discharge at 13 Pa for many hours. The generated plasma had a plasma potential of approximately 30 V and a relatively low plasma density ($5 \times 10^{15}$ m$^{-3}$), as measured with the Langmuir probe.

Figure 12 depicts the temperature curve upon exposure to the Ar plasma with ion kinetic energy of circa 30 eV. Heat flux was estimated to be 80 W/m$^2$ via the derivative of the fast transient upon switching plasma OFF (see sec. III.A). Under these plasma conditions, the sensor temperature readings stabilized at approximately 350 K two hours after plasma ignition and remained stable for more than eight hours of operation. A fast transient (see inset in Figure 12) followed by slow cooling was observed upon plasma extinction. Based on this and other multi-hour exposures, we can conclude that, when operating below or near the Ar RF plasma sputtering threshold[42,44], it does not significantly affect the lifetime and performance of a nanocalorimeter.

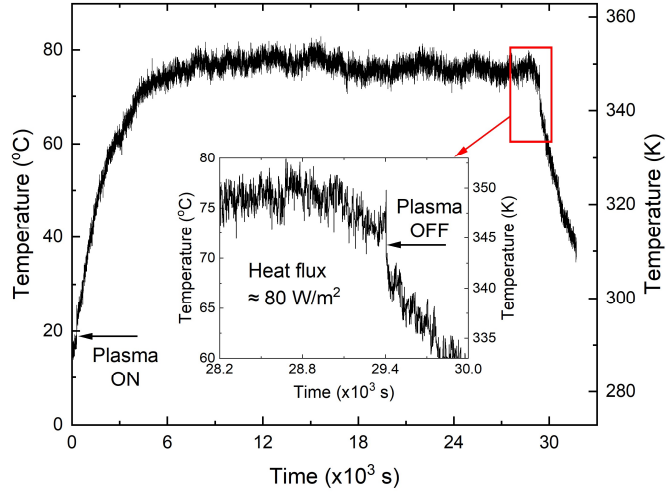

FIG. 12. Temperature evolution from nanocalorimeter 8 h longevity test. An RF discharge of Ar 13 Pa was held at 20 W in a CCP cell. Inset: heat flux was estimated from temperature decay when plasma was switched OFF.

The longevity tests above have been extended to the elevated bias conditions by repeating the previous ICP discharge heating experiments by ramping DC bias 10 V stepwise from 0 V up to -100 V (Figure 6a). Once the final value of -100 V is reached, the discharge is left operating until failure of the sensor due to ion etching. The destruction of the sensor is detected as the onset of the open circuit. Besides, physical breakage of the device is evident by visual inspection using scanning electron microscopy (SEM) (Figure 13). Note that the device membrane collapses by forming the characteristic roll-like structure (Figure 13b). The latter suggests the presence of the pre-existing strain in the device, which is common for layered and bimetallic structures[45]. The detailed investigation of membrane failure mechanisms is beyond the scope of this article and awaits a special study.

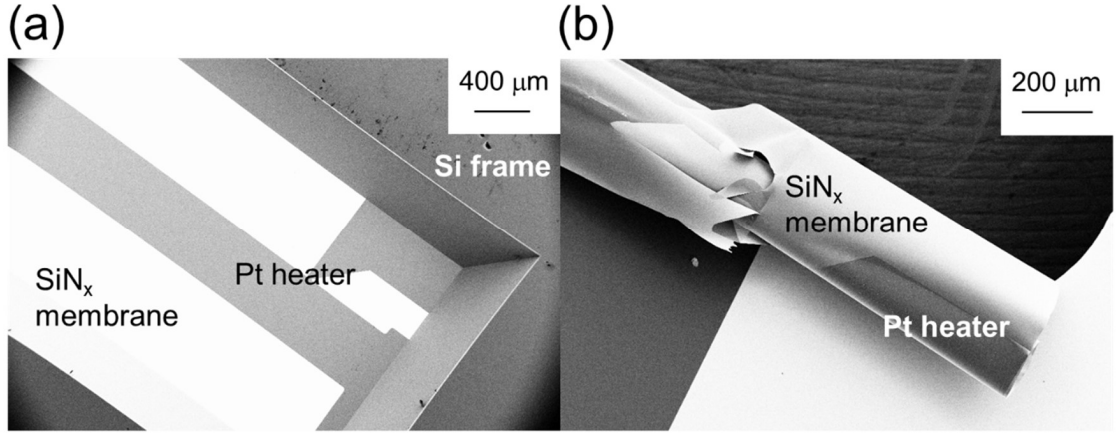

FIG. 13. Scanning electron microscopy (SEM) images of a nanocalorimeter (a) before and (b) after failure upon Ar plasma treatment at -100 V bias. The ruptured membrane appears rolled up at the edges of the Si frame.

The mean lifetimes due to energetic (130 eV) exposures with Ar and $O_2$ plasmas are approximately 15 min and 90 min, respectively. These experimental values match in order of magnitude with the predicted lifetimes using sputtering theory (Figure 11)[42].

## IV. CONCLUSIONS

The performance of a nanocalorimeter sensor based on a suspended thin-film Pt resistor has been tested in argon (inert) and oxygen (reactive) model plasmas generated by an RF ICP source. Inductive coupling enabled independent control over the plasma density and energy of plasma species, as well as the influence of these parameters on the sensor response. Two characteristic time domains are identified upon plasma exposure: (1) fast response associated with Pt sensor and membrane heat capacities, and (2) slow response determined by the Si frame heat capacity with some contribution from Al enclosure. The sensor (fast) response time to heat flux from plasma is on the order of 50 ms, thanks to the effective thermal decoupling from the supporting Si frame die. The sensor sensitivity, which is approximately 10 W/m² according to an unoptimized temperature detection limit of 5 K, is similar to that reported for other plasma thermal probes[32,33]. Thermal conduction to the gas and the Si frame is identified as the main heat sink under steady-state operation. The contributions of different plasma species on sensor heating have been decoupled and quantified by changing the DC bias on the sensor. Moreover, the use of DC bias has been aimed at enhancing the sensor's potential applicability in monitoring etching processes using energetic ions in typical plasma processing.

We show that positive plasma ions accelerated by negative bias become the dominant heating source above -50 V of DC bias, which corresponds to 80 eV in ion kinetic energy. Very high electron fluxes are collected when a positive bias voltage is applied in electropositive plasma, which may harm sensor functionality. On the contrary, this effect is drastically reduced in electronegative $O_2$ discharge, where ions dominate over electron current at positive bias on the sensor, thanks to its strong electronegativity. Overall, the kinetic energy and flux of ions pose limitations on sensor lifetime due to ion sputtering and the buildup of thermally driven stress in the membrane. The consideration of surface reaction cross sections is required to predict sensor stability within the more complex environments of reactive plasmas, such as fluorine- and chlorine-containing etching gases and their mixtures, which are typically employed in microfabrication facilities and necessitate a separate study.

This work has identified $j_i \cdot E_i$ as a key sensor heating channel by energetic plasma ions. This direct linkage between the energy of plasma ions and their thermal signature can be utilized in alternative plasma analyzers that rely on thermal input. Additionally, in view of the sensor stability, the present study suggests prospective uses of ultrathin $SiN_x$ membranes as differential pressure windows in emerging environmental-plasma XPS applications[46-48].

In summary, nanocalorimetry is a promising and sensitive metrology for non-thermal plasma environments, which are typical in semiconductor manufacturing. This technique provides a solution for obtaining sensitive and fast thermal measurements at the substrate position, and it nicely complements traditional plasma diagnostics, such as the Langmuir probe and RFEA. Moreover, the explored sensor's design is robust enough for continuous operation in reactive and non-reactive plasma discharges. Furthermore, via specifically designed and enhanced arrays, this technology can be integrated into microfabricated chips that collect and process electrical, optical, and thermal signals in a compact, all-in-one plasma diagnostics technique. Further device miniaturization will enable the characterization of plasma discharges with accurate spatial resolution, providing very fast measurements given the small dimensions of the involved sensors.

## SUPPLEMENTARY MATERIAL

A separate PDF file provides: (i) the temperature response of the nanocalorimeter exposed to Ar and $O_2$ plasmas, and (ii) a method to measure the sensor heat capacity.

# ACKNOWLEDGEMENTS

This work was funded by the CHIPS Metrology Program, part of CHIPS for America, National Institute of Standards and Technology, U.S. Department of Commerce. CHIPS for America has financially supported this work through the "Nanocalorimetry for Semiconductors and Semiconductor Process Metrology" project. Any mention of commercial products in this article is for information only; it does not imply recommendation or endorsement by NIST. The authors are thankful to Dr. Evgheni Strelcov, Dr. William A. Osborn, and Dr. Jason Campbell (all at NIST) for their careful reading of the manuscript and valuable suggestions. Mr. Dimitri Kolmakov's help in plasma source-chamber adaptor CAD designs is greatly appreciated.

# AUTHOR DECLARATIONS

**Conflicts of Interest**

The authors have no conflicts to disclose.

**Author Contributions**

**Carles Corbella:** Conceptualization (equal); Methodology (equal); Data curation (lead); Formal analysis (lead); Investigation (lead); Writing – original draft (lead). **Feng Yi:** Funding acquisition (lead); Project administration (equal); Resources (equal); Writing – review & editing (supporting). **Andrei Kolmakov:** Conceptualization (equal); Methodology (equal); Project administration (lead); Resources (equal); Software (supporting); Supervision (lead); Writing – review & editing (equal).

## DATA AVAILABILITY

The data that support the findings of this study are available from the corresponding author upon reasonable request.

# Supplementary Material for:

# Understanding microfabricated nanocalorimeter performance and responses to the energy fluxes from low-temperature plasma discharges

Carles Corbella[1,2], Feng Yi[1], and Andrei Kolmakov[3]

[1] Materials Measurement Science Division, MML, NIST, Gaithersburg, MD 20899, USA

[2] Department Chemistry & Biochemistry, University of Maryland, College Park, MD 20742, USA

[3] Nanoscale Device Characterization Division, PML, NIST, Gaithersburg, MD 20899, USA

## *A. Temperature evolution of argon and oxygen plasmas*

Figure S1 shows heating curves recorded with nanocalorimeter sensor upon exposure to Ar and $O_2$ RF ICP discharges operated at 6.5 Pa and 80 W. The sensor was electrically grounded. The higher heat flux measured for Ar discharge (640 W/m$^2$) compared to $O_2$ discharge (400 W/m$^2$) is correlated with the difference between argon ion flux (20 A/m$^2$) and oxygen ion flux (14 A/m$^2$) (see sections III.A.1 and III.B.1 in the main text). The long-term temperature drift behaviors in both discharges are similar, thereby suggesting comparable heat losses and effective heat capacities of the sensor.

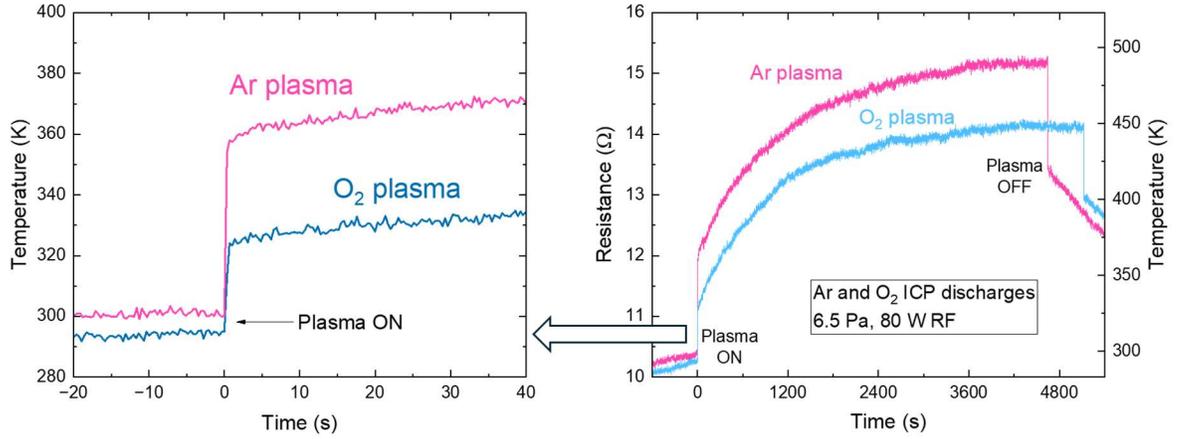

FIG. S1. Temporal evolutions of resistance and temperature of the nanocalorimeter (electrically grounded) receiving heat fluxes from Ar and $O_2$ plasmas. The expanded region shows two characteristic timescales of sensor heating upon plasma exposure.

## B. Experimental determination of sensor heat capacity

The heat capacity of nanocalorimeter sensor was measured via a thermal relaxation method based in Joule self-heating. We recorded the temperature transients due to switching Pt sensor current with 5 mA step (heating) and -5 mA step (cooling). Time derivative of temperature upon applying each current step, $\dot{T}$, and power variation equal to current × voltage product difference, $\Delta P$, were measured to obtain the sensor heat capacity, $C_s = \Delta P/\dot{T}$. Figure S2 shows two example temperature traces as a current step is applied. Figure S3 depicts heat capacity as a function of Ar pressure ($10^{-3}$ Pa to 270 Pa) and current variation at $10^{-3}$ Pa between 5 mA and 20 mA using 5 mA steps. An approximate $C_s \approx 2\times10^{-6}$ J/K was taken to evaluate energy flux from plasma in nanocalorimetry experiments.

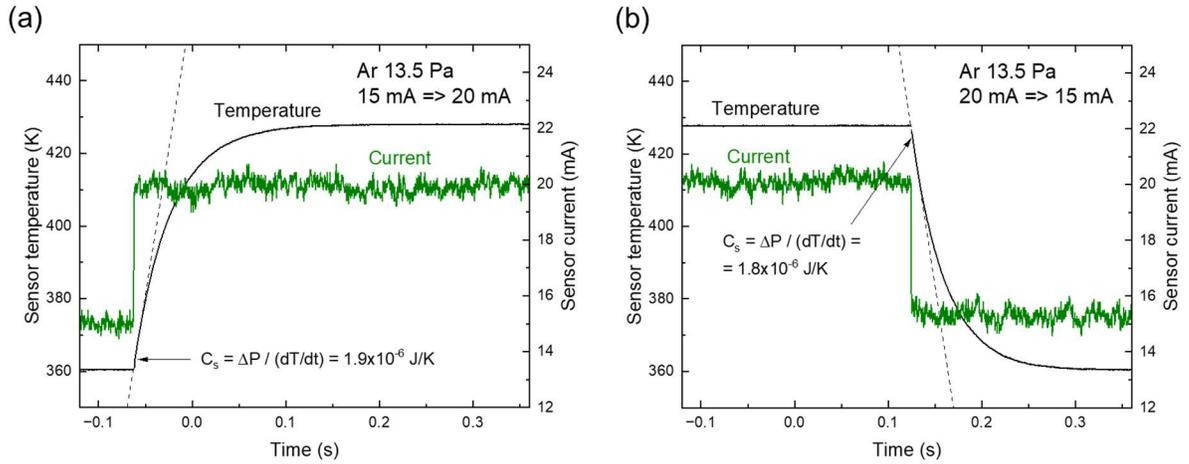

FIG. S2. Examples of heat capacity measurement via temperature relaxation. The curves of sensor temperature and current in Ar at 13.5 Pa were recorded for steps (a) 15 mA to 20 mA and (b) 20 mA to 15 mA.

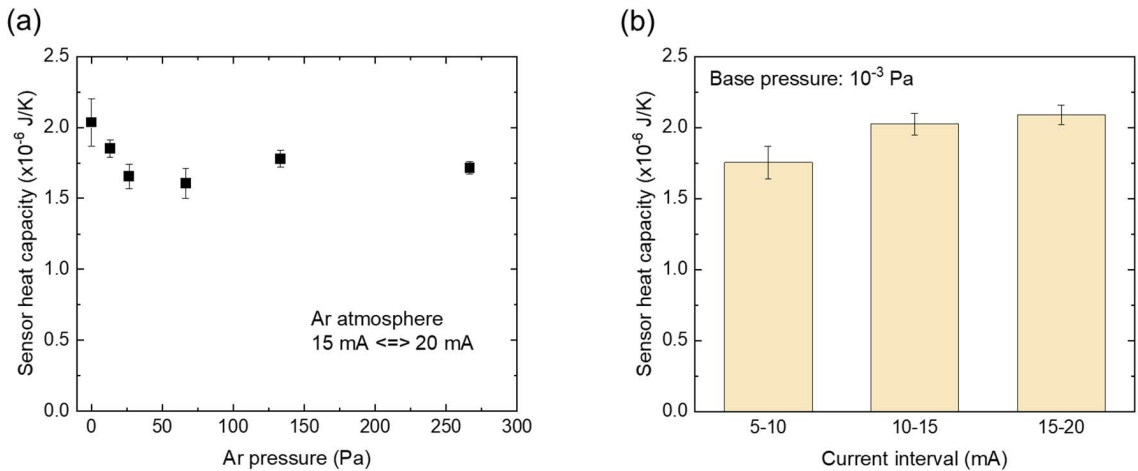

FIG. S3. Experimental values of sensor heat capacity as a function of (a) Ar pressure and (b) current step intervals. The error bars show the difference between heating- and cooling-step heat capacity measurements.